\documentclass[aps,nofootinbib,preprintnumbers,prd,notitlepage]{revtex4-1}

\usepackage{amsfonts}
\usepackage{amsmath}
\usepackage{braket}
\usepackage{dcolumn}
\usepackage{epstopdf}
\usepackage{graphicx}
\usepackage{hepparticles}
\usepackage{hepnicenames}
\usepackage{hepunits}
\usepackage{hyperref}
\usepackage{slashed}
\usepackage{subfigure}
\usepackage[svgnames]{xcolor}

\newcommand{\ie}{\textit{i.e.}}

\newcommand{\refapp}[1]{appendix~\ref{app:#1}}
\newcommand{\refeq}[1]{eq.~(\ref{eq:#1})}
\newcommand{\reffig}[1]{figure~\ref{fig:#1}}
\newcommand{\refsec}[1]{section~\ref{sec:#1}}
\newcommand{\reftab}[1]{table~\ref{tab:#1}}

\renewcommand{\theta}{\vartheta}
\newcommand{\Gfermi}{G_F}
\newcommand{\dd}[2][]{{\mathrm{d}^{#1}}#2\,}
\newcommand{\eps}{\varepsilon}

\let\Im\im
\let\Re\re

\newcommand{\qtau}{q_{[\tau]}}

\newcommand{\qmu}{q_{[\mu]}}
\newcommand{\qnubarmu}{q_{[\bar{\nu}_\mu]}}
\newcommand{\qnubartau}{q_{[\bar{\nu}_\tau]}}
\newcommand{\qnunubar}{q_{[\nu_\tau\bar\nu_\mu]}}

\newcommand{\thetamu}{\theta_{[\mu]}}
\newcommand{\thetatau}{\theta_{[\tau]}}
\newcommand{\thetast}{\theta^*_{[\mu]}}
\newcommand{\thetastst}{\theta^{**}_{[\bar\nu_\mu]}}
\newcommand{\phistst}{\phi^{**}}

\begin{document}

\allowdisplaybreaks

\preprint{EOS-2016-01,ZU-TH-2/16}
\title{Impact of leptonic $\tau$ decays on the distribution of $B\to P\mu\bar\nu$ decays}
\author{Marzia Bordone}
\email{mbordone@physik.uzh.ch}
\author{Gino Isidori}
\email{isidori@physik.uzh.ch}
\author{Danny van Dyk}
\email{dvandyk@physik.uzh.ch}
\affiliation{Physik Institut, University of Z\"urich, Winterthurerstrasse 190, CH-8057 Z\"urich, Switzerland}

\begin{abstract}
    We calculate the fully-differential rate of the decays $B\to P\tau(\to \mu\bar\nu\nu)\bar\nu$
    where $P = D,\pi$, which is a background to the semimuonic decays $B\to P\mu\bar\nu$. The decays
    with a $3\nu$ final state can have a sizeable impact on the experimental analyses of the ratios
    $R_D$ and $R_\pi$, depending on the event selection in the analysis. We outline a strategy
    which permits the extraction of $R_P \mathcal{B}(\tau \to \mu\bar\nu\nu)$ from the
    neutrino-inclusive rate. Our analytic results can also be used to test both existing and upcoming
    experimental analyses. We further provide Monte Carlo samples of the 5D rate of the neutrino-inclusive
    decays $B\to P\mu X_{\bar\nu}$.
\end{abstract}

\maketitle

\section{Introduction}
\label{sec:intro}

Charged-current semileptonic decays of $b$ hadrons are a precious source of
information about flavor physics, both within and beyond the Standard Model
(SM).  They are the primary source of information on the elements $|V_{cb}|$
and $|V_{ub}|$ of the Cabibbo-Kobayashi-Maskawa (CKM) mixing
matrix~\cite{Cabibbo:1963yz,Kobayashi:1973fv,Kowalewski:2014:PDG} and, at the
same time, they offer the possibility of interesting tests of physics beyond
the SM via appropriate Lepton Flavor Universality (LFU) ratios.  In this paper
we concentrate on the simplest of such LFU ratios, namely
\begin{equation}
    R_P = \frac{\mathcal{B}(\bar{B}\to P\tau\bar\nu)}{\mathcal{B}(\bar{B}\to P\mu\bar\nu)}\,,
\end{equation}
 where $P=D,\pi$.

The theoretical estimate of $ R_P$ within the SM relies dominantly on the
hadronic form factors $f_+$ (the vector form factor) and $f_0$ (the scalar form
factor), see \refapp{form-factors} for their definitions. For both final
states, precise lattice QCD result of these form factors have recently been
published \cite{Lattice:2015tia,Na:2015kha}. In addition, Light-Cone Sum Rules
(LCSRs) results for the $B\to \pi$ vector form factor and two of its
derivatives have been obtained, which complement the lattice QCD results.
According to these studies the SM prediction for $R_D$ \cite{Na:2015kha} is
\begin{equation}
    R_D^\text{SM} = 0.300 \pm 0.008\,.
\end{equation}
On the experimental side, measurements of the ratio $R_D$ have been published by both BaBar
\cite{Lees:2013uzd} and, more recently, by Belle \cite{Huschle:2015rga},
\begin{equation}
\begin{aligned}
    R_D^\text{BaBar} & = 0.440 \pm 0.058 \pm 0.042\,, &
    R_D^\text{Belle} & = 0.375 \pm 0.064 \pm 0.026\,,
\end{aligned}
\end{equation}
while only upper experimental bounds on $R_\pi$ are available \cite{Hamer:2015jsa}.
Combining Babar and Belle results, and normalizing them to the SM, leads to
\begin{equation}
    \Delta R_{D}  =  \frac{R_D^{\rm exp}}{R_D^\text{SM}} - 1 = 0.35 \pm 0.17\,.
\end{equation}
This deviation from the SM is not particularly significant; however, a similar
effect has been observed also in the $R_{D^*}$
ratios~\cite{Lees:2013uzd,Huschle:2015rga,Aaij:2015yra}.  Combining the two
deviations, which are compatible with a universal enhancement of semileptonic
$b\to c\tau\nu$ transitions over $b\to c\mu\nu$ ones, the discrepancy with
respect to the SM raises to about $\sim 4\sigma$. This fact has stimulated
several studies on possible New Physics (NP) explanations (see
e.g.~Ref.~\cite{Celis:2012dk,Alonso:2015sja,Greljo:2015mma,Calibbi:2015kma}).  As pointed out in
Ref.~\cite{Greljo:2015mma}, because of  $\tau \to \ell  \bar\nu \nu $ decays, a
possible enhancement of semileptonic $b\to c\tau\nu$ transitions may have a
non-trivial impact in the extraction of $|V_{cb}|$  from the corresponding
$b\to c\ell \nu$ modes, and this impact is likely to be different  for
exclusive and inclusive modes.

Our main goal is to analyze how leptonic $\tau \to \mu \bar\nu \nu$ decays
affect the determination of $R_P$ and, more generally, the kinematical
distribution of $\bar{B}\to P \mu\bar\nu $ decays via the decay chain
$\bar{B}\to P\tau(\to \mu\bar\nu \nu)\bar\nu$ in experimental analyses  where there 
is no precise information available on the missing mass (or the initial $B$ momentum).
As we will discuss, our results
provide a first attempt toward new  strategies to improve the determination of
$R_P$ from data and, possibly, also the determination of $|V_{cb}|$ and
$|V_{ub}|$.  At first glance, leptonic $\tau$ decay modes might seem
unimportant, since they occur at the expense of an additional power of the
Fermi coupling $\Gfermi$ at the amplitude level.  However, this process occurs
on-shell and the suppression of the $\tau$ decay amplitude is compensated by
the inverse of the $\tau$ lifetime appearing in the $\tau$ propagator. 
This becomes already apparent in the $\tau \to \mu\bar{\nu}_\mu \nu_\tau$ branching
fraction: $\mathcal{B}(\tau \to \mu\bar\nu_\mu \nu_\tau) = (17.41 \pm 0.04)\%$ \cite{Agashe:2014kda}.
It is therefore interesting to calculate the rate for the decay chain $\bar{B}\to
P\tau(\to \mu\bar\nu \nu)\bar\nu$, and compute numerically its impact on the
observable rate of $\bar{B}\to P\mu X_{\bar\nu}$, $X_{\bar\nu} = \lbrace
\bar\nu, \bar\nu\nu\bar\nu\rbrace$, to which we will henceforth refer as the
``neutrino-inclusive'' decay.

The layout of this article is as follows. We continue in \refsec{setup} with
definitions and the bulk of our analytical results. Numerical results and their
implications are presented in \refsec{numerics}, and we summarize in
\refsec{summary}. The appendices contain details on the form factors in
\refapp{form-factors}, details on the kinematic variables in
\refapp{scalar-products}, and the numeric results of the $3\nu$ PDFs in
\refapp{pdf-results}.

\section{Setup}
\label{sec:setup}

\subsection{Kinematics}

As anticipated in the introduction, in this article we assume that experiments
cannot distinguish between  the semileptonic decay $\bar{B}\to P\mu\bar\nu$ and
$\bar{B}\to P\tau(\to \mu\bar\nu\nu)\bar\nu$ using the missing-mass
information.  This assumption certainly holds for analyses performed at hadron
colliders (e.g., by the LHCb experiment\footnote{See the supplementary material to
ref.~\cite{Aaij:2015yra}, figure 9.}).  On the other hand, it does not hold
for analyses performed at $e^+e^-$  colliders with flavour tagging based on the
full reconstruction of the opposite $B$ decay, where  $\bar{B}\to P\mu\bar\nu$
and $\bar{B}\to P\tau(\to \mu\bar\nu\nu)\bar\nu$ will be clearly distinguished
using the missing-mass information.  The latter type of analyses will certainly
provide precise results in the future; however, they  cannot be performed at
present and will require high statistics.  It is therefore useful to discuss
the case where there is no (or poor) missing-mass information.\\

We write for the neutrino-inclusive differential decay width to one muon:
\begin{equation}
\begin{aligned}
    \frac{\dd\Gamma(\bar{B}\to P \mu X_{\bar\nu})}{\dd{q^2} \dd{\!\cos\thetamu}}
    & \equiv \frac{\dd \Gamma(\bar{B}\to P \mu \bar\nu_\mu)}{\dd{q^2} \dd{\!\cos\thetamu}}
    + \frac{\dd \Gamma(\bar{B}\to P \tau (\to \mu \bar\nu_\mu \nu_\tau) \bar\nu_\tau)}{\dd{q^2} \dd{\!\cos\thetamu}}\\
    & \equiv \frac{\dd \Gamma_1}{\dd{q^2} \dd{\!\cos\thetamu}}
    + \frac{\dd \Gamma_3}{\dd{q^2} \dd{\!\cos\thetamu}}\,.
\end{aligned}
\end{equation}
In the above, we introduce the shorthand $\Gamma_n$ for the specific decay width with $n=1$ or $n=3$
neutrinos in the final state.\footnote{%
    We also drop the subscript for the neutrino flavor where possible. Note that effects of
    neutrino mixing and/or oscillation are not relevant to our study.
}
The kinematic variable are defined as follows.
\begin{itemize}
    \item We define $q^\mu$ as the momentum transfer away from the $\bar{B}$-$P$ system, \ie: $q^\mu \equiv p^\mu - k^\mu$,
        where $p$ and $k$ are the momenta of the $\bar{B}$ and $P=D,\pi$ mesons, respectively.
        For $\Gamma_1$ this implies that $q^\mu$ coincides with the momentum of the lepton pair $\mu\bar\nu_\mu$.
        We stress that this does not hold for $\Gamma_3$.
    \item We define the angle $\thetamu$ via
        \begin{equation}
            \cos\thetamu \equiv 2\frac{\left(q - 2q_{[\mu]}\right) \cdot k}{\sqrt{\lambda}}\,.
        \end{equation}
        We abbreviate the K\"all\'en function $\lambda \equiv \lambda(M_B^2, M_P^2, q^2)$ here and throughout
        this article. For $\Gamma_1$, the above formula coincides with
        \begin{equation}
            \cos\thetamu = 2\frac{\left(q_{[\bar\nu_\mu]} - q_{[\mu]}\right)\cdot k}{\sqrt{\lambda}}\,,
        \end{equation}
        and the physical meaning of $\thetamu$ is the helicity angle of the muon in the $\mu\bar\nu_\mu$ rest frame,
        with $-1 \leq \cos\theta_\mu \leq +1$. We stress that for $\Gamma_3$ this physical interpretation is \emph{no longer valid}.
        Yet, we find it convenient to keep using $\cos\thetamu$ for the description of the neutrino-inclusive rate $\Gamma(\bar{B}\to P \mu X_{\bar\nu})$.
        We emphasize also that the phase space boundaries for $\cos\thetamu$ in $\Gamma_3$ differ from those in $\Gamma_1$,
        and implicitly depend on the full kinematics of the $3\nu$ decays.
\end{itemize}
For the description of $\Gamma_3$, we need to define further kinematic variables, which will be integrated over
at a later point. We choose $q_{[\tau]}^2$, the mass square of the $\tau$ lepton; $q_{[\nu_\tau \bar\nu_{\mu}]}^2 \equiv (q_{[\nu_\tau]} + q_{[\bar\nu_{\mu}]})^2$,
the mass square of the two neutrinos produced in the $\tau$ decay; as well as five angles:
\begin{enumerate}
    \item $\thetatau$, the helicity angle of the $\tau$ in the $\tau\bar\nu_\tau$ rest frame:
\begin{equation}
    \cos\thetatau = \frac{(q - 2 \qtau)\cdot k}{\beta_\tau \sqrt{\lambda}} + \frac{(1 - 2\beta_\tau)}{\beta_\tau} \frac{(M_B^2 - M_P^2 - q^2)}{2 \sqrt{\lambda}}\,,
\end{equation}
    where $2 \beta_\tau \equiv 1 - \qtau^2 / q^2$,
    \item $\phi$, the azimuthal angle between the $\mu$-$\nu_\tau\bar\nu_\mu$ plane and the $\bar{B}$-$\tau\bar\nu_\tau$ plane,
\begin{equation}
    \eps(p, q, \qmu, \qnunubar) = -\frac{1}{2} \beta_{\nu\bar\nu} \sqrt{1 - 2 \beta_\tau} \beta_\tau q^2 \sqrt{\lambda} \sin\phi \sin \thetast \sin \thetatau\,,
\end{equation}
    \item $\thetast$, the polar angle of the $\mu$ momentum in the $\tau$ rest frame with respect to $\qnunubar$
        in the $\tau$ rest frame:
\begin{equation}
    \cos\thetast = \frac{1}{2\beta_{\nu\bar\nu} \beta_\tau} \left[(1 - 2\beta_{\nu\bar\nu}) (1 - \beta_\tau) + \frac{(\qmu - \qnunubar) \cdot q}{q^2}\right]\,,
\end{equation}
    where $2 \beta_{\nu\bar\nu} \equiv 1 - \qnunubar^2 / \qtau^2$,
    \item $\thetastst$, the polar angle of the $\bar\nu_\mu$ momentum in the $\nu_\tau\bar\nu_\mu$ rest frame
        with respect to the $\mu$ momentum in the $\nu_\tau\bar\nu_\mu$ rest frame:
\begin{equation}
    \cos\thetastst = \frac{(\qnunubar - 2 \qnubarmu) \cdot \qmu}{\beta_{\nu\bar\nu} \qtau^2}\,.
\end{equation}
    \item $\phistst$, the azimuthal angle between the $\tau$-$\mu$ and $\bar\nu_\mu$-$\nu_\tau$ decay planes in the $\tau$ rest frame,
\begin{equation}
    \eps(\qtau, \qnubartau, \qmu, \qnunubar) = \frac{1}{2} \beta_{\nu\bar\nu} \beta_\tau \sqrt{1 - 2 \beta_{\nu\bar\nu}} q^2 \qtau^2 \sin\thetast \sin\thetastst \sin\phistst\,.
\end{equation}
\end{enumerate}
In general, we denote the solid angle in the $\tau\bar\nu_\tau$ rest frame without any asterisks, the solid angle within the $\tau$ rest
frame with one asterisk, and the solid angle in the $\bar\nu_\mu\nu_\tau$ rest frame with two asterisks.\\

With the above definitions of the kinematics in mind, we can now begin
discussing phenomenological applications. We wish to first address the case, in
which a $3\nu$ event is misinterpreted as a 1-neutrino event. In such a
case, the misreconstructed $\cos\thetamu$ reads
\begin{multline}
    \label{eq:zmu}
    \cos\thetamu\Big|_{3\nu}
    = 2 \beta_{\nu\bar\nu} \Bigg\lbrace
        \left(\frac{(1 - 2\beta_{\nu\bar\nu})}{\beta_{\nu\bar\nu}} + 2 \beta_\tau \right)\frac{M_B^2 - M_P^2 - q^2}{2 \sqrt{\lambda}}
        + \beta_\tau \cos\thetatau\\
        - \left( 2\beta_\tau \frac{M_B^2 - M_P^2 - q^2}{2 \sqrt{\lambda}} - (1 - \beta_\tau) \cos\thetatau \right) \cos\thetast
        - \sqrt{1 - 2 \beta_\tau} \sin\thetast \sin\thetatau \cos\phi
    \Bigg\rbrace\,.
\end{multline}
As an alternative to $\cos\thetamu$ we also consider $E_\mu$, the muon energy in the
$B$ rest frame. It is defined in terms of Lorentz invariants as
\begin{equation}
    E_\mu \equiv \frac{p \cdot \qmu}{M_B}\,.
\end{equation}
In the $1\nu$ decay, $E_\mu$ is not independent from our nominal choice of kinematic variables
$q^2$ and $\cos\theta_\mu$. The expression for $E_\mu$ reads
\begin{equation}
    E_\mu\Big|_{1\nu}
    = \frac{1}{4 M_B}\left[(M_B^2 - M_P^2 + q^2) - \sqrt{\lambda} \cos\theta_\mu\right]\,,
\end{equation}
and it attains its maximal value at $q^2 = 0$ and $\cos\theta_\mu = -1$. Its full range reads
\begin{equation}
    m_\mu \leq E_\mu\Big|_{1\nu} \leq \frac{M_B^2 - M_P^2}{2 M_B}\,.
\end{equation}
However, for a misreconstructed $3\nu$ event we obtain instead
\begin{multline}
    \label{eq:Emu}
    E_\mu\Big|_{3\nu}
    = \frac{\beta_{\nu\bar\nu}}{2 M_B}\Big[(M_B^2 - M_P^2 + q^2)((1 - \beta_\tau) + \beta_\tau \cos\thetast)\\
    - \sqrt\lambda (\beta_\tau + (1 - \beta_\tau) \cos\thetast)\cos\thetatau + \sqrt{1 - 2 \beta_\tau} \sqrt\lambda \sin\thetast \sin\thetatau \cos\phi\Big]\,,
\end{multline}
which now exhibits an additional dependence on the kinematics variables $\cos\thetast$ and $\phi$, as well as $\qnunubar^2$.
We find for its range
\begin{equation}
    m_\mu \leq E_\mu\Big|_{3\nu} \leq \frac{M_B^2 - M_P^2 + m_\tau^2 + \sqrt{\lambda(M_B^2, M_P^2, m_\tau^2)}}{4 M_B}\,.
\end{equation}

\subsection{Decay Rate}

In order to proceed, we require an analytic expression for the
neutrino-inclusive differential decay rate.  The result for $\Gamma_1$ is known
for some time in the literature (see e.g.~\cite{Dutta:2013qaa,Becirevic:2016hea} for
reviews in the presence of model-independent NP contributions). However,
$\Gamma_3$ has not been calculated to the best of our knowledge. We begin the
computation with the matrix element for the $\bar{B}(p) \to P(k)
\tau(q_{[\tau]}) \bar\nu(q_{[\bar\nu_\tau]})$ transition:
\begin{equation}
    i \mathcal{M} = -i \frac{\Gfermi V_{cb}}{\sqrt{2}} \left[ f_+(q^2) \left\lbrace (p + k)^\mu - \frac{M_B^2 - M_P^2}{q^2} q^\mu \right\rbrace
                    + f_0(q^2) \frac{M_B^2 - M_P^2}{q^2} q^\mu\right] L^{(V-A)}_\mu\,,
\end{equation}
with $q \equiv p - k = q_{[\tau]} + q_{[\bar\nu_\tau]}$. In the above, we abreviate the leptonic currents as
\begin{equation}
    \begin{aligned}
        L_\mu^{(V-A)} & \equiv \left[\bar{u}(q_{[\tau]}) \gamma_\mu (1 - \gamma_5) v(q_{[\bar\nu_\tau]})\right]\,.
    \end{aligned}
\end{equation}
The contributions to $\Gamma_3$ then arise from the leptonic decay of the $\tau$. The corresponding matrix
elements can be readily obtained through the replacement
\begin{equation}
    \begin{aligned}
        L_\mu^{(V-A)} & \mapsto \frac{-i \Gfermi}{\sqrt{2}} \frac{i}{\qtau^2 - m_\tau^2 + i m_\tau \Gamma_\tau} \tilde{L}_\mu^{(V-A)}\\
                      & = \frac{\Gfermi}{\sqrt{2}(q_{[\tau]}^2 - m_\tau^2 + i m_\tau \Gamma_\tau)} \left[\bar{u}(q_{[\mu]}) \gamma_\alpha(1-\gamma_5) v(q_{[\bar\nu_\mu]})\right]\\
                      & \quad \times \left[\bar{u}(q_{[\nu_\tau]}) \gamma^\alpha (1 - \gamma_5) (\slashed{q}_{[\tau]} + m_\tau)\gamma_\mu (1 - \gamma_5) v(q_{[\bar\nu_\tau]})\right]\,,\\
    \end{aligned}
\end{equation}
where $m_\tau$ and $\Gamma_\tau$ denote the mass and the total width of the $\tau$ lepton, respectively.

The fully-differential rate for the 3-neutrino final state can then be expressed as:
\begin{multline}
    \label{eq:d7Gamma}
    \frac{\dd[7]{\Gamma_3}}{\dd{q^2}\dd{\qnunubar^2}\dd[2]{\Omega} \dd{\Omega^*} \dd[2]{\Omega^{**}}}
    = -\frac{3 \Gfermi^2 |V_{cb}|^2 \sqrt{\lambda} (q^2 - m_\tau^2) (m_\tau^2 - \qnunubar^2) \mathcal{B}(\tau \to \mu\bar\nu\nu)}{2^{17} \pi^5 m_\tau^8 M_B^3 q^2}\\
    \times\Big[
        |f_+|^2 \left(T_1 - \frac{M_B^2 - M_D^2}{q^2} T_2 + \frac{(M_B^2 - M_D^2)^2}{q^4} T_3\right)\\
        + \Re{(f_+\,f_0)} \left(\frac{M_B^2 - M_D^2}{q^2} T_2 - 2 \frac{(M_B^2 - M_D^2)^2}{q^4} T_3\right)
        + |f_0|^2 \frac{(M_B^2 - M_D^2)^2}{q^4} T_3
    \Big]\,,
\end{multline}
with auxilliary quantities
\begin{equation}
\begin{aligned}
    T_1 & \equiv (p + k)^\mu (p + k)^\nu \sum_\text{spins} \tilde{L}^{(V-A)}_\mu \tilde{L}^{*,(V-A)}_\nu\,,\\
    T_2 & \equiv \left((p + k)^\mu q^\nu + (p + k)^\nu q^\mu\right) \sum_\text{spins} \tilde{L}^{(V-A)}_\mu \tilde{L}^{*,(V-A)}_\nu\,,\\
    T_3 & \equiv q^\mu q^\nu \sum_\text{spins} \tilde{L}^{(V-A)}_\mu \tilde{L}^{*,(V-A)}_\nu\,.
\end{aligned}
\end{equation}
In the above we abbreviate $\dd[2]{\Omega} = \dd{\!\cos\thetatau} \dd{\phi}$, $\dd{\Omega^*} = \dd{\!\cos\thetast}$, and
$\dd[2]{\Omega^{**}} = \dd{\!\cos\thetastst} \dd{\phistst}$, and we emphasize that the integration range over $\dd{\cos \theta}$
goes from from $-1$ to $+1$. The full expressions for $T_{1,2,3}$ are quite cumbersome to typeset. Instead, we opt to publish
them as ancillary files within the arXiv preprint of this article.
We also find that the integration of \refeq{d7Gamma} over $\Omega^{**}$, $\Omega^*$, $\phi$ and $\qnunubar^2$ yields
$\mathcal{B}(\tau \to \mu\bar\nu_\mu \nu_\tau) \times \dd[2]{\Gamma(\bar{B}\to P\mu\bar\nu)}/\dd{q^2} \dd{\!\cos\thetatau}$
as required. This is a successful crosscheck of our calculation.

In order to carry out our phenomenological study of the quantities $\cos\thetamu$ in \refeq{zmu}
and $E_\mu$ in \refeq{Emu} in the decay chain $\bar{B}\to P\tau(\to \mu \bar{\nu}\nu)\bar\nu$,
we do not require any dependence on the $\nu\bar\nu$ solid angle $\Omega^{**} = (\cos\thetastst, \phistst)$.
We therefore integrate over the latter, and thus obtain the five-differential rate
\begin{equation}
    \label{eq:d5Gamma}
    \frac{\dd[5]{\Gamma_3}}{\dd{q^2}\dd{\qnunubar^2}\dd[2]{\Omega} \dd{\Omega^*}}
    = \frac{\tilde\Gamma_3}{\pi m_\tau^8 q^6} \left[A + B \cos\thetatau + C \cos^2\thetatau
    + \left(D \sin\thetatau + E \sin\thetatau \cos\thetatau\right) \cos\phi\right]\,,
\end{equation}
with normalization
\begin{equation}
    \tilde\Gamma_3 = \frac{|V_{cb}|^2\Gfermi^2 \mathcal{B}(\tau\to \mu\nu\bar\nu)}{2^{9} \pi^3 M_B^3}\,.
\end{equation}
The angular coefficients in \refeq{d5Gamma} read
\begin{equation}
\begin{aligned}
    A & = [(q^2 - m_\tau^2) (m_\tau^2 - \qnunubar^2)]^2 \sqrt{\lambda} \Big[(m_\tau^2 + 2 \qnunubar^2) (|f_0|^2 (M_B^2 - M_P^2)^2 m_\tau^2 + |f_+|^2 q^2 \lambda)
\\
      & \qquad - (m_\tau^2 - 2 \qnunubar^2) (|f_0|^2 (M_B^2 - M_P^2)^2 m_\tau^2 - |f_+|^2 q^2 \lambda) \cos\thetast\Big]\,,\\
    B & =  2|f_0||f_+|m_\tau^2(M_B^2 - M_P^2)\lambda[(q^2 - m_\tau^2) (m_\tau^2 - \qnunubar^2)]^2\Big[(m_\tau^2 + 2 \qnunubar^2)-(m_\tau^2 - 2 \qnunubar^2) \cos\thetast\Big]\,,\\
    C & = -|f_+|^2\lambda^{3/2}[(q^2 - m_\tau^2) (m_\tau^2 - \qnunubar^2)]^2\Big[(q^2 - m_\tau^2)(m_\tau^2 + 2 \qnunubar^2) + (q^2 + m_\tau^2)(m_\tau^2 - 2 \qnunubar^2)\cos\thetast\Big]\,,\\
    D & = 2 m_\tau \sqrt{q^2}|f_0||f_+|(M_B^2 - M_P^2)[(q^2 - m_\tau^2)(m_\tau^2 - \qnunubar^2)]^2(m_\tau^2 - 2 \qnunubar^2)\lambda\sin\thetast\,,\\
    E & = 2 m_\tau \sqrt{q^2}|f_+|^2[(q^2 - m_\tau^2)(m_\tau^2 - \qnunubar^2)]^2(m_\tau^2 - 2 \qnunubar^2)\lambda^{3/2}\sin\thetast\,.
\end{aligned}
\end{equation}
We can now proceed to to produce the pseudo-events that are distributed as \refeq{d5Gamma}, which is a
necessary prerequisite for our phenomenological applications in the following section.

\section{Numerical results}
\label{sec:numerics}

Our numerical results are based on a Monte Carlo (MC) study of the decays $\bar{B}\to P \mu \bar\nu$
and $\bar{B}\to P \tau(\to \mu\nu\bar\nu) \bar\nu$. For this purpose, we added the signal PDFs for
both decays to the EOS library of flavor observables \cite{EOS}. The relevant form factors $f_+$ and
$f_0$ are taken in the BCL parametrization \cite{Bourrely:2008za}. The BCL parameters are
fitted from a recent lattice QCD studies \cite{Na:2015kha, Lattice:2015tia}, and additionally Light-Cone Sum Rules
results in the case of $\bar{B}\to \pi$ \cite{Imsong:2014oqa}; see \refapp{form-factors} for details.

In order to obtain pseudo events for the neutrino inclusive decay, we carry out the following steps:
\begin{enumerate}
    \item We draw $4.8\cdot 10^6$ samples $\lbrace\vec{X}^{(1)}_i\rbrace = \lbrace (q^2, \cos\thetamu)_i \rbrace$,
        which are distributed as their signal PDF $P_1$,
        \begin{equation}
            P_1(q^2, \cos\thetamu) \equiv \frac{1}{\Gamma_1} \frac{\dd[2]{\Gamma_1}}{\dd{q^2} \dd{\cos\thetamu}}\,.
        \end{equation}
    \item We draw $4.8\cdot 10^6$ samples $\lbrace\vec{X}^{(3)}_i\rbrace = \lbrace (q^2, \qnunubar^2, \cos\thetatau, \phi, \cos\thetast)_i \rbrace$,
        which are distributed as their signal PDF $P_3$,
        \begin{equation}
            P_3(q^2, \qnunubar^2, \cos\thetatau, \phi, \cos\thetast) \equiv \frac{1}{\Gamma_3} \frac{\dd[5]{\Gamma_3}}{\dd{q^2} \dd{\qnunubar^2} \dd{\cos\thetatau} \dd{\phi} \dd{\cos\thetast}}\,.
        \end{equation}
    \item We combine the two sets of samples with weights $\omega_1 = \Gamma_1 / (\Gamma_1 + \Gamma_3)$
        and $\omega_3 = 1 - \omega_1$, respectively. The weights can be expressed in terms
        of $R_P$ and $\mathcal{B}(\tau \to \mu\nu\bar\nu)$:
        \begin{equation}
            \omega_1 = \frac{1}{1 + R_P \mathcal{B}(\tau \to \mu\nu\bar\nu)}\,.
        \end{equation}
\end{enumerate}
All samples are obtained from a Markov Chain Monte Carlo setup, which
implements the Metropolis-Hastings algorithm
\cite{Metropolis:1953am,Hastings:1970aa}. The first $8\cdot 10^5$ samples per
set are discarded, in order to minimize the impact from the Markov Chains'
starting values. In order to avoid correlations from rejection of proposals, we
only take every tenth sample. The effective sample size is therefore $4\cdot 10^5$.
We provide the so-obtained pseudo events online \cite{EOS-DATA-2016-01} in the
binary HDF5 format\footnote{See \url{https://www.hdfgroup.org/HDF5/} for its
description.}.

\subsection{$\bar{B}\to D \mu X_{\bar\nu}$}

\begin{figure}
    \subfigure[%
        \label{fig:predictions-BtoDXnu-zmu}
    ]
    {%
        \includegraphics[width=.32\textwidth]{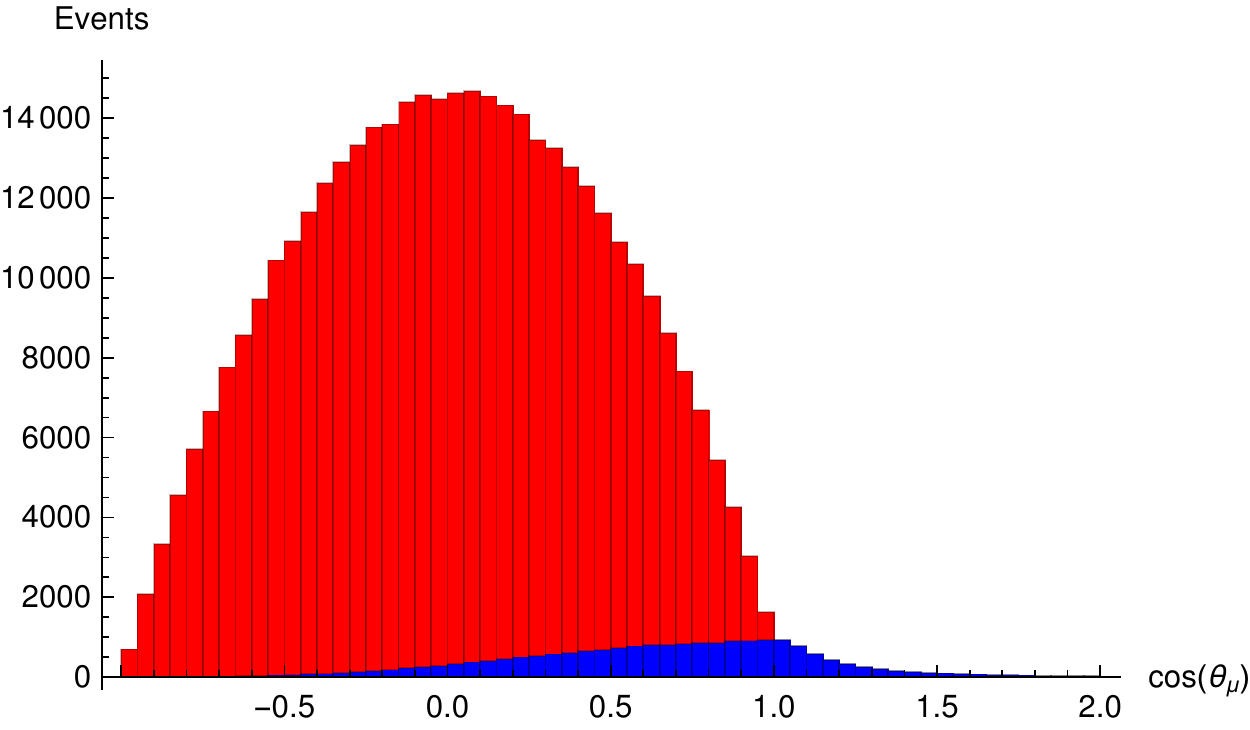}
    }
    \subfigure[%
        \label{fig:predictions-BtoDXnu-emu}
    ]
    {%
        \includegraphics[width=.32\textwidth]{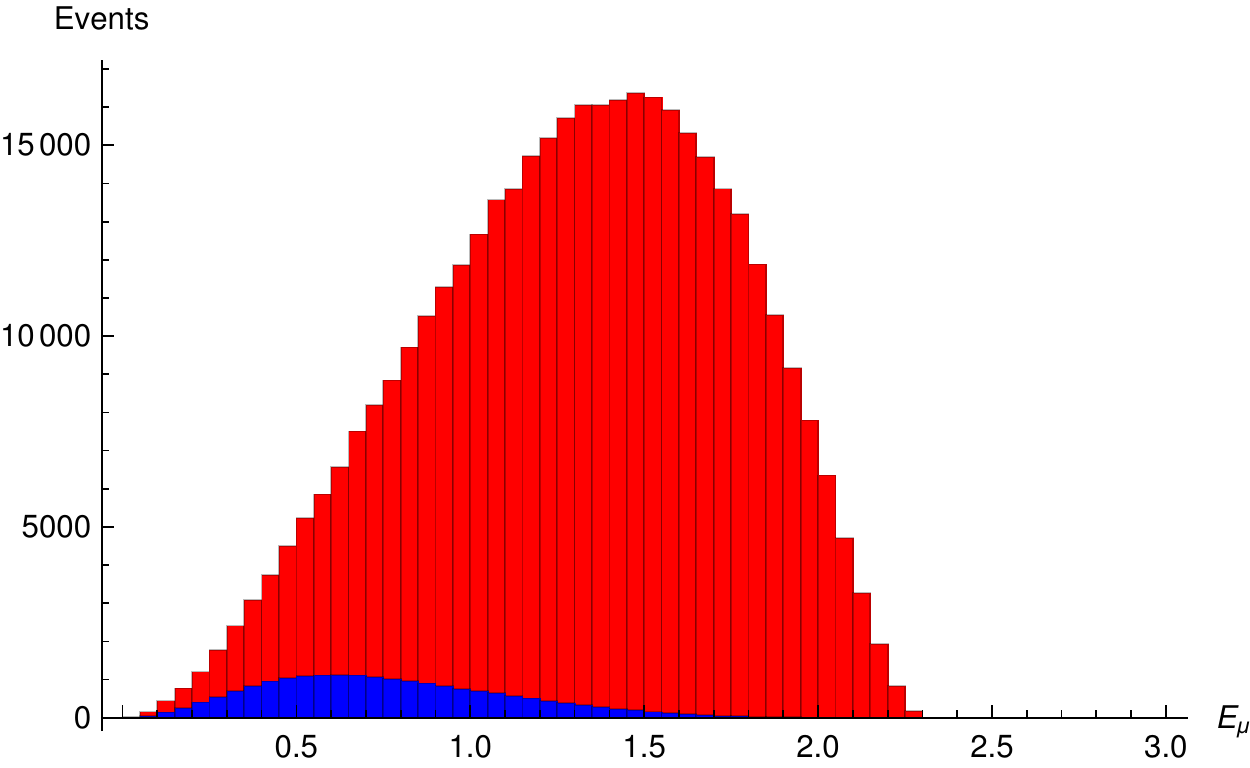}
    }
    \subfigure[%
        \label{fig:predictions-BtoD3nu-emu-Py}
    ]
    {%
        \includegraphics[width=.32\textwidth]{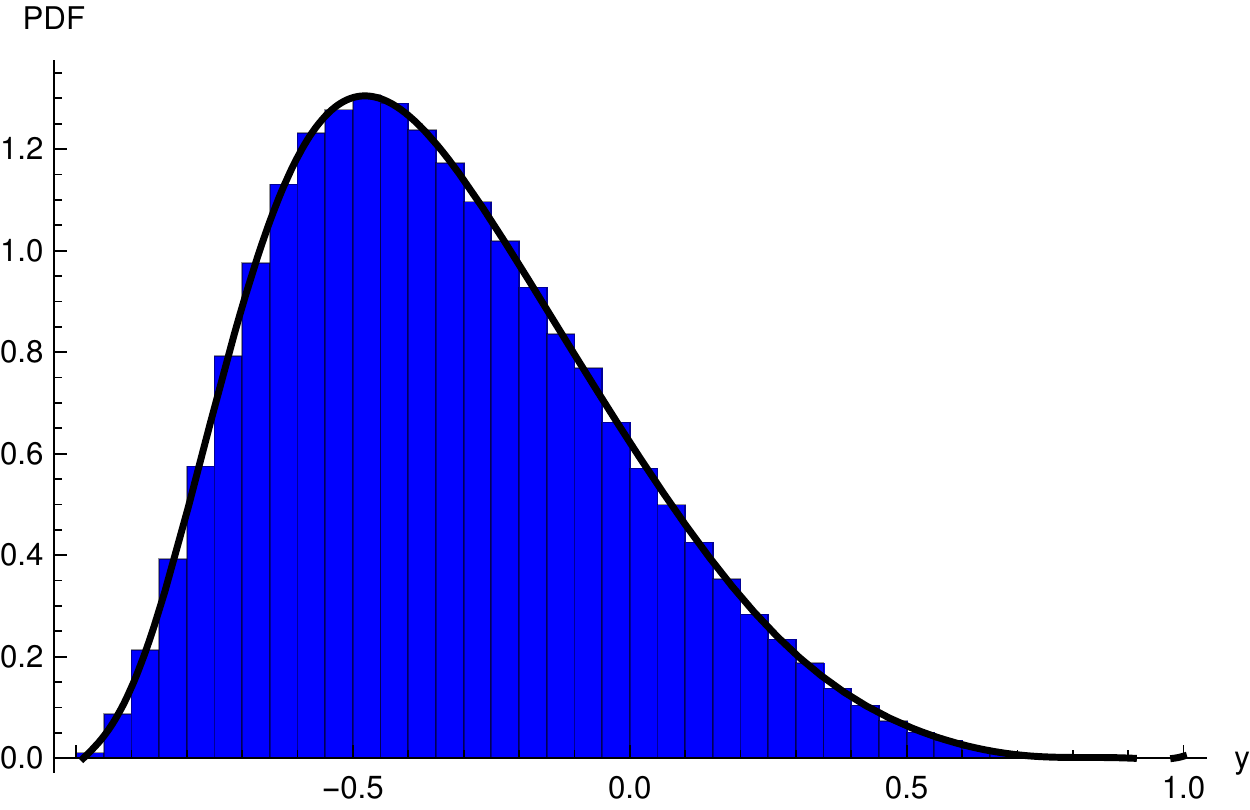}
    }
    \caption{
        Histograms of $4\cdot 10^5$ pseudo events for the neutrino inclusive decay
        $B \to D \mu X_{\bar\nu}$ [figures (a) and (b)], as well as for the decay
        $B \to D \tau(\to \mu\nu\bar\nu)\bar\nu$ [figure (c)].
        We show histograms of distributions in the (misreconstructed) angle $\cos\theta_{\mu}$
        [figure (a)], and $E_\mu$, the muon energy in the $\bar{B}$ rest frame [figure (b)].
        The red areas correspond to the neutrino-inclusive decay, while the
        blue areas highlight the contributions stemming only from $B\to D\tau(\to \mu\nu\bar\nu)\bar\nu$.
        We also show the histogram of $E_\mu\big|_{3\nu}$ and its compatibility with our ansatz
        \refeq{ansatz-Py} [figure (c)].
    }
\end{figure}

\paragraph*{Distribution in $\cos\thetamu$} In the neutrino inclusive decay, the
misreconstructed observable $\cos\thetamu$ as given in \refeq{zmu} is no
longer bounded by $+1$. We find that it attains its maximal value
\begin{equation}
    \max \cos\thetamu\big|_{3\nu} \simeq 56.7\qquad\text{for } q^2 = (M_B - M_D)^2,\quad\qnunubar^2 = m_\tau^2,\quad\cos\thetatau = -\cos\thetast = 1\,.
\end{equation}
The distribution of $\cos\thetamu$ in the neutrino-inclusive decay is shown
in \reffig{predictions-BtoDXnu-zmu}, where we also disentangle the individual $1\nu$
and $3\nu$ contributions. We find that $\cos\thetamu$ exceeds $1$ for
$\sim 23\%$ of the $3\nu$ events, and exceeds $2$ for $\sim 1.3\%$ of $3\nu$ events.
As a consequence, we decide against a parametrization of the neutrino-inclusive
PDF $P(\cos\theta_\mu)$ in terms of Legendre polynomials (or any other orthonormal
polynomial basis).

On the other hand, our findings imply that the $\cos\thetamu$ distribution can be used to extract 
the product $ R_D \mathcal{B}(\tau \to \mu\bar\nu\nu)$ from data.
We can indeed write 
\begin{equation}
    R_D \mathcal{B}(\tau \to \mu\bar\nu\nu) = \frac{\rho^{\rm exp}_{D}}{\rho_D^0 - \rho^{\rm exp}_{D}}
\end{equation}
where 
\begin{equation}
    \rho_D^0 \equiv \frac{\# \text{of $3\nu$ events with } \cos\theta_\mu > 1}{\text{total \# of $3\nu$ events}}\,, \qquad 
    \rho^{\rm exp}_{D}   \equiv \frac{\# \text{of $X_\nu$ events with } \cos\theta_\mu > 1}{\text{total \# of $X_\nu$ events}}
\end{equation}
Based on our MC pseudo events, we find
\begin{equation}
    \label{eq:ND3nu}
    \rho^0_D = 0.234 \pm 0.001\,
\end{equation}
where the error is dominantly statistical, and arises from our limited number of MC samples.
We explicitly cross check our uncertainty estimate by re-running the simulations with modified
inputs on the $B\to D$ form factors. We find that shifting any single individual constraint in
\reftab{lattice-points-BtoD} by $1\sigma$ yields results that are compatible with the interval
given in \refeq{ND3nu}.\\

\paragraph*{The distribution in $E_\mu$}

The distribution of $E_\mu$ in the neutrino-inclusive decay is shown in
\reffig{predictions-BtoDXnu-emu}. We find that a lower cut $E_\mu > 1.0\,\GeV$ can reduce
the rate of of misidentified $3\nu$ events by a factor of $\sim 4$, while $\sim 76\%$ of the
$1\nu$ events (the signal) remain. This corresponds
to a reduction of the rate of background events in the neutrino-inclusive decay from its maximum value
of $R_D \mathcal{B}(\tau \to \mu\nu\bar\nu) \approx 5.2\%$ down to $1.3\%$.

Alternatively, one can subtract the $3\nu$ background from the neutrino-inclusive rate. For this purpose
we proceed to obtain the relevant PDF of $3\nu$ events. Since the ranges of $E_\mu\Big|_{1\nu}$ and
$E_\mu\Big|_{3\nu}$ are very similar, we can remap their union to a new kinematic variable $y$,
\begin{equation}
    y \equiv \frac{2 E_\mu}{E_\mu^\text{max}} - 1\,,
    \qquad\text{with }E_\mu^\text{max} = \max\left(E_\mu\Big|_{1\nu},\,E_\mu\Big|_{3\nu}\right) \simeq 2.31\,\GeV\,,
    \qquad\text{so that } -1 \leq y \leq +1\,.
\end{equation}
We then make an ansatz for the PDF $P_3(y) \equiv \dd{\Gamma_3}/\dd{y}$ by expanding in
Legendre polynomials $p_k(y)$:
\begin{equation}
    \label{eq:ansatz-Py}
    P_3(y) = \frac{1}{2} + \sum_{k=1}^{12} c^{(3)}_k p_k(y)\,.
\end{equation}
Since the Legendre polynomials form an orthogonal basis of function on the support $[-1, +1]$, the
coefficients $c^{(3)}_k$ are independent of the degree of $P_3(y)$. Their mean values and covariance
are obtained using the method of moments; see \cite{Beaujean:2015xea} for a recent review.
We find that our ansatz \refeq{ansatz-Py} describes the PDF exceptionally well, and refer to
\reffig{predictions-BtoD3nu-emu-Py} for the visualization. Our results for the mean values and covariance
matrix of the moments are compiled in \reftab{results-legendre-moments-BtoD}. They can be used in
upcoming experimental studies in order to cross check the signal/background discrimination.

\subsection{$\bar{B}\to \pi \mu X_{\bar\nu}$}

\begin{figure}
    \subfigure[%
        \label{fig:predictions-BtopiXnu-zmu}
    ]
    {%
        \includegraphics[width=.32\textwidth]{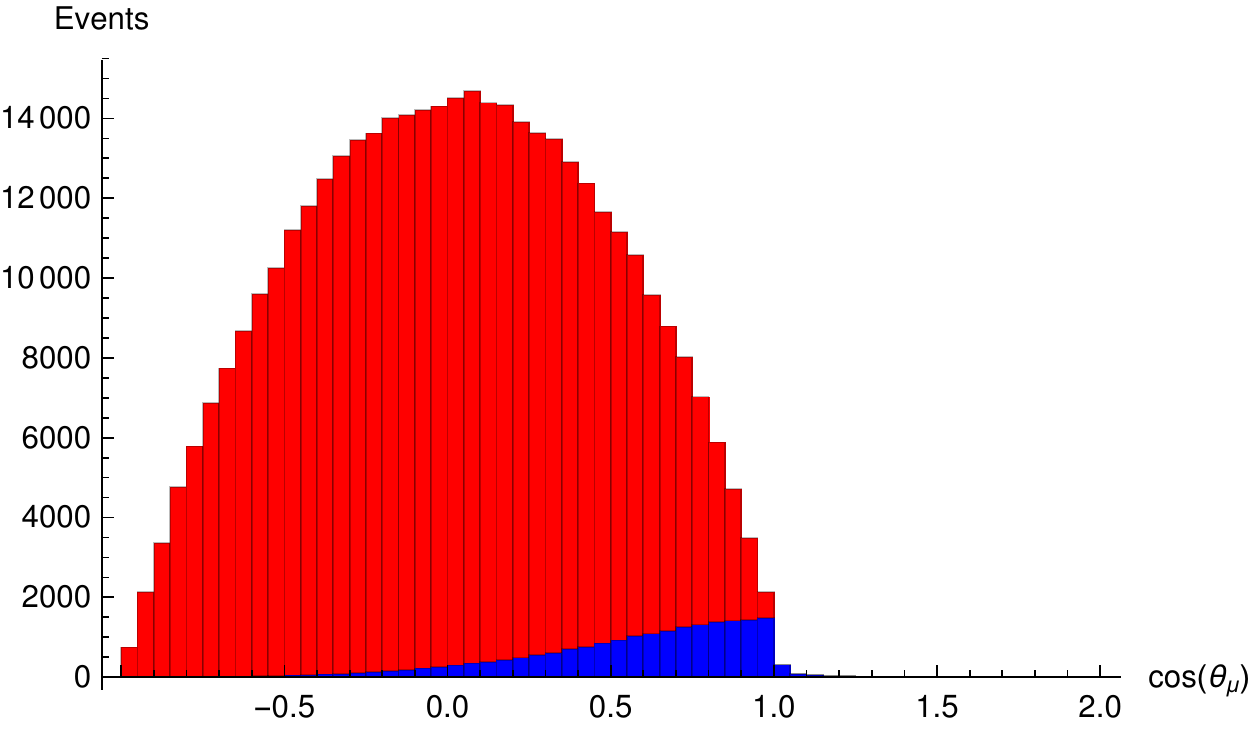}
    }
    \subfigure[%
        \label{fig:predictions-BtopiXnu-emu}
    ]
    {%
        \includegraphics[width=.32\textwidth]{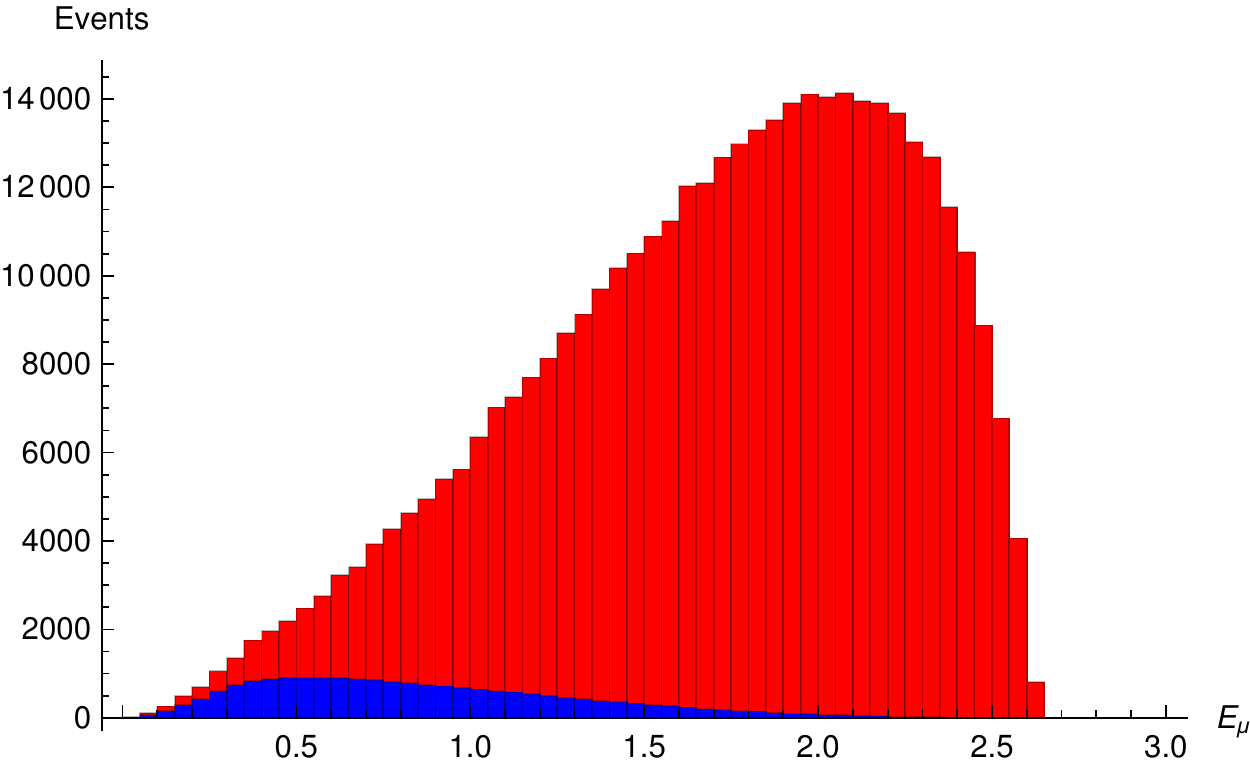}
    }
    \subfigure[%
        \label{fig:predictions-Btopi3nu-emu-Py}
    ]
    {%
        \includegraphics[width=.32\textwidth]{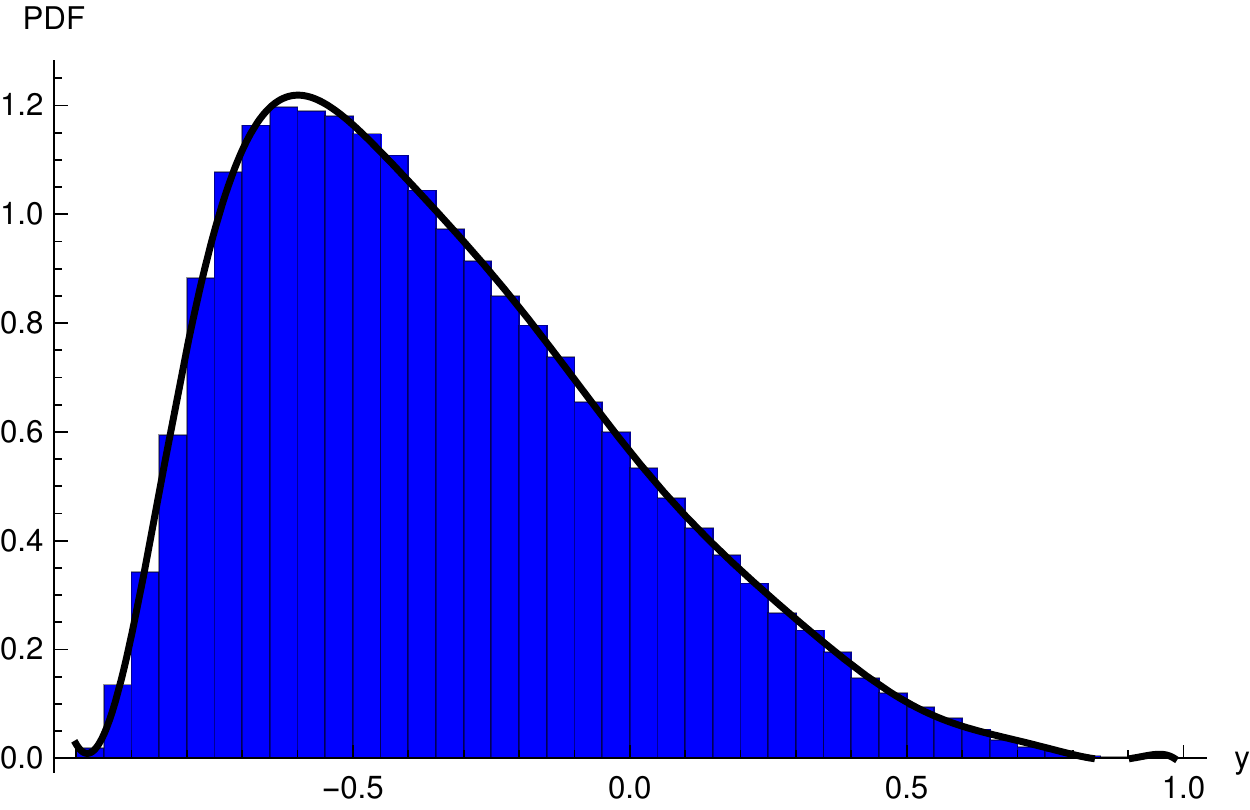}
    }
    \caption{
        Histograms of $4\cdot 10^5$ pseudo events for the neutrino inclusive decay
        $B \to \pi \mu X_{\bar\nu}$ [figures (a) and (b)], as well as for the decay
        $B \to \pi \tau(\to \mu\nu\bar\nu)\bar\nu$ [figure (c)].
        We show histograms of distributions in the (misreconstructed) angle $\cos\theta_{\mu}$
        [figure (a)], and $E_\mu$, the muon energy in the $\bar{B}$ rest frame [figure (b)].
        The red areas correspond to the neutrino-inclusive decay, while the
        blue areas highlight the contributions stemming only from $B\to \pi\tau(\to \mu\nu\bar\nu)\bar\nu$.
        We also show the histogram of $E_\mu\big|_{3\nu}$ and its compatibility with our ansatz
        \refeq{ansatz-Py} [figure (c)].
    }
\end{figure}

Based on the $\bar{B}\to \pi$ form factors parameters as described in \refapp{form-factors}, we obtain
\begin{equation}
    R_\pi^\text{SM} = 0.70 \pm 0.01\,,
\end{equation}
which is in good visual agreement with the plot of $R_\pi$ in figure 8 of Ref.~\cite{Khodjamirian:2011ub}.
This result implies a potentially larger impact of the $3\nu$ decays as a background in the extraction
of both $R_\pi$ and $|V_{ub}|$.\\

\paragraph*{Distribution in $\cos\thetamu$} As in the case of $\bar{B}\to D\mu X_{\bar\nu}$, the misreconstructed
observable $\cos\thetamu$ is no longer bounded from above by $+1$. However, we find that its maximal value
is much smaller for $\bar{B}\to \pi$ transitions than it is for $\bar{B}\to D$ transitions:
\begin{equation}
    \max \cos\thetamu\big|_{3\nu} \simeq 3.75\qquad\text{for } q^2 = (M_B - M_\pi)^2,\quad\qnunubar^2 = m_\tau^2,\quad\cos\thetatau = -\cos\thetast = 1\,.
\end{equation}
A consequence of this smaller upper bound in $\bar{B}\to \pi$ transitions, the
tail of $3\nu$ events is much lighter; see \reffig{predictions-BtopiXnu-zmu}. This is also reflected
in our numerical result for the ratio $\rho^0_\pi$,
\begin{equation}
    \label{eq:Npi3nu}
    \rho^0_{\pi} = (2.89 \pm 0.03) \cdot 10^{-2}\,.
\end{equation}
We can therefore not recommend to extract the ratio $R_\pi$ through a lower cut on $\cos\thetamu$.
Our result also shows that more than $97\%$ of $3\nu$ events fall in the physical region of $1\nu$ events.\\

\paragraph*{Distribution in $E_\mu$} We find that a lower cut $E_\mu > 1.5\,\GeV$ can reduce
the rate of of misidentified $3\nu$ events by a factor of $\sim 10$, while $\sim 69\%$ of the
$1\nu$ events (the signal) remain. This corresponds
to a reduction of the rate of background events in the neutrino-inclusive decay from its maximum value
of $R_\pi \mathcal{B}(\tau \to \mu\nu\bar\nu) \simeq 12.1\%$ down to $\sim 1.2\%$.

For the range of $E_\mu$ we find
\begin{equation}
    \max \left(E_\mu\Big|_{1\nu}\,, E_\mu\Big|_{3\nu}\right) \simeq 2.64\,\GeV\,,
\end{equation}
and the energy ranges are overlapping given our numerical precision. Thus, the description of the neutrino-inclusive
rate though $E_\mu$, or equivalently $y$, should work even better for $\bar{B}\to \pi$ transitions
than for $\bar{B}\to D$ transitions. Our results for the mean values and covariance matrix of the
Legendre moments $c_k^{(3)}$ are compiled in \reftab{results-legendre-moments-Btopi}. We refer to \reffig{predictions-Btopi3nu-emu-Py}
for a comparison of $P_3(y)$ with our MC pseudo events.

\subsection{Implications for the extraction of $|V_{cb}|$ and $|V_{ub}|$}

Using the above results we can finally draw some semi-quantitative conclusions about the error in
the extraction $|V_{cb}|$ and $|V_{ub}|$ from $b\to c(u) \ell \nu$ decays. The presence of the
$\tau \to \mu\bar\nu\nu$ background in those processes can be dealt with, experimentally, in different ways.
The two extreme cases we can envisage are the following: i) reduction of the background via explicit cuts;
ii) fully inclusive subtraction. The first method can be applied to exclusive decays such as those discussed
in the present paper. As shown above, combining cuts in $E_\mu$ and $\cos\thetamu$ leads to a significant 
reduction of the $\tau \to \mu\bar\nu\nu$ contamination in $\bar{B}\to D\mu X_{\bar\nu}$, with
negligible implications for the extraction of $|V_{cb}|$. However, this procedure cannot be applied to fully inclusive modes.
In the latter case, the $\tau \to \mu\bar\nu\nu$  contamination is more likely to be simply subtracted from
the total number of events. If this subtraction is made assuming the SM expectation of $R_D$ (and $R_{D*}$), it
leads to systematic error if $\Delta R_{D} \not =0$, i.e.~in presence of New Physics~\cite{Greljo:2015mma}.
The maximal value of this error is
\begin{equation}
\frac{  \Delta |V_{cb}|^{(\rm incl.)}  }{ |V_{cb}| }  = \frac{1}{2} \Delta R_{D}   \mathcal{B}(\tau \to \mu\bar\nu\nu)   \approx 0.9 \%\,,
\end{equation}
which is not far from the combined theory and experimental error presently quoted for $|V_{cb}|$ \cite{Agashe:2014kda}.
We thus conclude that  the  $\tau \to \mu\bar\nu\nu$ contamination must be carefully analyzed
in the determination of $|V_{cb}|$.

The impact of the $\tau \to \mu\bar\nu\nu$ contamination is more difficult to
be estimated in the $|V_{ub}|$ case. On the one hand, the large value of
$R_\pi$ leads to a potentially larger impact. On the other hand, even in
inclusive analyses some cut on $E_\mu$ is unavoidable in order to reduce the
$b\to c \ell \nu$ background: as shown above, this naturally leads to a
significant reduction of the $\tau \to \mu\bar\nu\nu$ contamination. Given the
present large experimental errors, the $\tau \to \mu\bar\nu\nu$ contamination
is likely to be a subleading correction in the extraction of $|V_{ub}|$, but it
is certainly an effect that has to be properly  analyzed in view of future
high-statistics data.

\section{Summary}
\label{sec:summary}

Lepton Flavor Universality tests in charged-current semileptonic $B$ decays provide 
a very interesting window on possible physics beyond the SM.
In the paper we have  analyzed  how the leptonic $\tau \to \mu \bar\nu \nu$ decays
affect the determination of the LFU ratios $R_P$, where $P=D,\pi$. 
In particular, we have presented a complete analytical determination 
of the observable distributions (energy spectrum and helicity angle of the muon) 
of the $\bar{B}\to P\tau(\to \mu\bar\nu \nu)\bar\nu$ decay chain.
This result has allowed us to identify clean strategies both to extract $R_P$ from 
measurements of the $\bar{B}\to P\mu X_{\bar\nu}$ neutrino-inclusive rate,
and also to minimize the impact of the $\tau \to \mu \bar\nu \nu$ decay 
in the three-body $\bar{B}\to P \mu \nu$ modes. 
Finally, this study has also allowed us to conclude that the $b \to c  \tau (\to \ell \bar\nu \nu) \nu $ 
background in $b\to  c \ell \nu$ decays represents a non-negligible source
of uncertainty for the extraction of $|V_{cb}|$
in presence of NP modifying $R_D$: its impact could reach the $\sim1\%$ level and  
has to be analyzed with care mode by mode.

\acknowledgments

We thank Heechang Na for useful communications on the lattice QCD analysis in \cite{Na:2015kha}.
We gratefully acknowledge discussions with Thomas Kuhr about semileptonic
analyses at Belle and Belle-II, and with Nicola Serra on semileptonic analyses
at LHCb. D.v.D. also thanks Frederik Beaujean for helpful discussions.\\

This research was supported in
part by the Swiss National Science Foundation (SNF) under contract 200021-159720
and contract PP00P2-144674.

\appendix

\section{$\bar{B}\to P$ form factors}
\label{app:form-factors}

\begin{table}
    \renewcommand{\arraystretch}{1.4}
    \begin{tabular}{l|ccccccc}
        \text{}           & $f_+(0\,\GeV^2)$       &  $f_+(4\,\GeV^2)$       &  $f_+(8\,\GeV^2)$       & $f_+(t_-)$              &  $f_0(4\,\GeV^2)$       & $f_0(8\,\GeV^2)$        &  $f_0(t_-)$            \\
    \hline
        \multicolumn{8}{c}{mean}\\
    \hline
        \text{}           & $0.665$                &  $0.798$                &  $0.972$                &  $1.177$                &  $0.729$                &  $0.810$                &  $0.901$               \\
    \hline
        \multicolumn{8}{c}{covariance matrix}\\
    \hline
        $f_+(0\,\GeV^2)$  & $1.128\times 10^{-3}$  &  $1.042\times 10^{-3}$  &  $9.230\times 10^{-4}$  &  $7.727\times 10^{-4}$  &  $1.093\times 10^{-3}$  &  $1.063\times 10^{-3}$  &  $1.045\times 10^{-3}$ \\
        $f_+(4\,\GeV^2)$  & $1.042\times 10^{-3}$  &  $1.079\times 10^{-3}$  &  $1.108\times 10^{-3}$  &  $1.123\times 10^{-3}$  &  $1.026\times 10^{-3}$  &  $1.017\times 10^{-3}$  &  $1.021\times 10^{-3}$ \\
        $f_+(8\,\GeV^2)$  & $9.230\times 10^{-4}$  &  $1.108\times 10^{-3}$  &  $1.331\times 10^{-3}$  &  $1.576\times 10^{-3}$  &  $9.307\times 10^{-4}$  &  $9.511\times 10^{-4}$  &  $9.865\times 10^{-4}$ \\
        $f_+(t_-)$        & $7.727\times 10^{-4}$  &  $1.123\times 10^{-3}$  &  $1.576\times 10^{-3}$  &  $2.112\times 10^{-3}$  &  $8.108\times 10^{-4}$  &  $8.681\times 10^{-4}$  &  $9.425\times 10^{-4}$ \\
        $f_0(4\,\GeV^2)$  & $1.093\times 10^{-3}$  &  $1.026\times 10^{-3}$  &  $9.307\times 10^{-4}$  &  $8.108\times 10^{-4}$  &  $1.126\times 10^{-3}$  &  $1.165\times 10^{-3}$  &  $1.210\times 10^{-3}$ \\
        $f_0(8\,\GeV^2)$  & $1.063\times 10^{-3}$  &  $1.017\times 10^{-3}$  &  $9.511\times 10^{-4}$  &  $8.681\times 10^{-4}$  &  $1.165\times 10^{-3}$  &  $1.283\times 10^{-3}$  &  $1.410\times 10^{-3}$ \\
        $f_0(t_-)$        & $1.045\times 10^{-3}$  &  $1.021\times 10^{-3}$  &  $9.865\times 10^{-4}$  &  $9.425\times 10^{-4}$  &  $1.210\times 10^{-3}$  &  $1.410\times 10^{-3}$  &  $1.635\times 10^{-3}$ \\
    \end{tabular}
    \renewcommand{\arraystretch}{1.0}
    \caption{Mean values and covariance matrix for the data points reconstructed from \cite{Na:2015kha} at $q^2 \in \lbrace 0\,\GeV^2, 4\,\GeV^2, 8\,\GeV^2, t_- = (M_B^2 - M_D^2)\rbrace$.}
    \label{tab:lattice-points-BtoD}
\end{table}

\begin{table}
    \renewcommand{\arraystretch}{1.4}
    \begin{tabular}{l|cccccc}
        \text{}           & $f_+(18\,\GeV^2)$     & $f_+(22\,\GeV^2)$     & $f_+(26\,\GeV^2)$     & $f_0(18\,\GeV^2)$     & $f_0(22\,\GeV^2)$     & $f_0(26\,\GeV^2)$     \\
    \hline
        \multicolumn{7}{c}{mean}\\
    \hline
        \text{}           & $1.016$               & $1.971$               & $6.443$               & $0.417$               & $0.609$               & $0.961$               \\
    \hline
        \multicolumn{7}{c}{covariance matrix}\\
    \hline
        $f_+(18\,\GeV^2)$ & $3.492\times 10^{-3}$ & $1.997\times 10^{-3}$ & $1.648\times 10^{-3}$ & $1.067\times 10^{-3}$ & $2.904\times 10^{-4}$ & $1.096\times 10^{-4}$ \\
        $f_+(22\,\GeV^2)$ & $1.997\times 10^{-3}$ & $3.371\times 10^{-3}$ & $6.193\times 10^{-3}$ & $2.123\times 10^{-4}$ & $2.167\times 10^{-4}$ & $1.294\times 10^{-4}$ \\
        $f_+(26\,\GeV^2)$ & $1.648\times 10^{-3}$ & $6.193\times 10^{-3}$ & $7.419\times 10^{-2}$ & $2.064\times 10^{-3}$ & $1.139\times 10^{-3}$ & $1.346\times 10^{-3}$ \\
        $f_0(18\,\GeV^2)$ & $1.067\times 10^{-3}$ & $2.123\times 10^{-4}$ & $2.064\times 10^{-3}$ & $8.478\times 10^{-4}$ & $4.266\times 10^{-4}$ & $3.150\times 10^{-4}$ \\
        $f_0(22\,\GeV^2)$ & $2.904\times 10^{-4}$ & $2.167\times 10^{-4}$ & $1.139\times 10^{-3}$ & $4.266\times 10^{-4}$ & $3.923\times 10^{-4}$ & $4.009\times 10^{-4}$ \\
        $f_0(26\,\GeV^2)$ & $1.096\times 10^{-4}$ & $1.294\times 10^{-4}$ & $1.346\times 10^{-3}$ & $3.150\times 10^{-4}$ & $4.009\times 10^{-4}$ & $6.467\times 10^{-4}$ \\
    \end{tabular}
    \renewcommand{\arraystretch}{1.0}
    \caption{Mean values and covariance matrix for the data points reconstructed from \cite{Lattice:2015tia} at $q^2 \in \lbrace 18\,\GeV^2, 22\,\GeV^2, 26\,\GeV^2\rbrace$.}
    \label{tab:lattice-points-Btopi}
\end{table}

The hadronic matrix element for the vector current between two pseudoscalar states
is commonly (e.g.~\cite{Bourrely:2008za}) expressed in terms of two form factor
\begin{equation}
    \label{eq:def-form-factors}
    \bra{P(k)} \bar{c} \gamma^\mu b \ket{\bar{B}(p)}
        = f_+(q^2) \left[(p + k)^\mu - \frac{M_B^2 - M_P^2}{q^2} q^\mu\right]
        + f_0(q^2) \frac{M_B^2 - M_P^2}{q^2} q^\mu\,.
\end{equation}
In the above, $q^\mu \equiv p^\mu - k^\mu$. In the limit $q^2 \to 0$ one finds
a relation between the two form factors in the form of
\begin{equation}
    \label{eq:eom-form-factors}
    f_+(0) = f_0(0)\,,
\end{equation}
otherwise \refeq{def-form-factors} would diverge.

While the heavy quark limit can be used as a guiding principle to parametrize
both form factors, we prefer not to apply it. Instead, we follow the BCL ansatz
\cite{Bourrely:2008za} and write
\begin{equation}
    \label{eq:param-form-factors}
\begin{aligned}
    f_+(q^2) & = \frac{f_+(0)}{1 - q^2/M_{R(1^-)}^2} \left[1 + \sum_{k=1}^3 \alpha^+_k z^k(q^2; t_+, 0)\right]\,,\\
    f_0(q^2) & = \frac{f_+(0)}{1 - q^2/M_{R(0^+)}^2} \left[1 + \sum_{k=1}^2 \alpha^0_k z^k(q^2, t_+, 0)\right]\,,
\end{aligned}
\end{equation}
where $M_{R(1^-)}$ and $M_{R(0^+)}$ denote the masses of the low-lying resonances with
spin/parity quantum numbers $J=1^-$ and $J=0^+$, respectively.
Note the use of $f_+(0)$
in the parametrization of $f_0(q^2)$, which automatically fulfills the equation of
motion \refeq{eom-form-factors}. In the parametrization \refeq{param-form-factors},
we make use of the conformal mapping from $q^2$ to $z$, where
\begin{equation}
    z(q^2; t_+, t_0) = \frac{\sqrt{t_+ - q^2} - \sqrt{t_+ - t_0^{\phantom{2}}}}{\sqrt{t_+ - q^2} - \sqrt{t_+ - t_0^{\phantom{2}}}}\,.
\end{equation}
Following \cite{Bourrely:2008za} we impose $\Im{f_{+}(q^2)} = (q^2 - t_+)^{3/2}$ close to
the pair-production threshold $t_+ \equiv (M_B + M_D)^2$. This leads to a relation
between the expansion parameters $\alpha^+_k$:
\begin{equation}
    \alpha^+_3 = \frac{1}{3} \sum_{k=1}^{K-1} (-1)^k k\, \alpha^+_k\,.
\end{equation}

\paragraph*{$\bar{B}\to D$} The lattice QCD results as presented in
\cite{Na:2015kha} follow the BCL parametrization, however, they do not
automatically fulfill the equation of motion \refeq{eom-form-factors}.  We
therefore reconstruct lattice data points for four different choices of $q^2$
(see \reftab{lattice-points-BtoD}), and fit our choice of the parametrization
to these reconstructed points. We use $M_{R(1^-)} = 6.330\GeV$ and $M_{R(0^+)}
= 6.420\GeV$ as in \cite{Na:2015kha}.

\paragraph*{$\bar{B}\to \pi$} The lattice QCD results as presented in
\cite{Lattice:2015tia} follow the BCL parametrization.  However, they do not
automatically fulfill the equation of motion \refeq{eom-form-factors}.
Moreover, for the form factor $f_0(q^2)$, no pole for a low-lying resonance
scalar resonance is used. We therefore reconstruct lattice data points for
three different choices of $q^2$ in the domain for which lattice data point had
been obtained (see \reftab{lattice-points-Btopi}).  In addition, we use the
results of a recent Light-Cone Sum Rules (LCSR) study \cite{Imsong:2014oqa} for
the form factor $f_+$ at $q^2 = \lbrace 0, 10\rbrace\,\GeV^2$. The LCSR results
provide, beyond the form factor $f_+$, also its first and second derivatives
with respect to $q^2$.  We fit our choice of the parametrization to the
aforementioned constraints. We use $M_{R(1^-)} = 5.325\GeV$ and $M_{R(0^+)} =
5.540\GeV$.

\section{Scalar Products}
\label{app:scalar-products}

In order to facilitate the comparison with our results, we list here all
scalar products that emerge in the calculation of \refeq{d7Gamma}.\\

The scalar products involving $p$ are
\begin{align}
    p \cdot q     & = \frac{M_B^2 + q^2 - M_D^2}{2}\,,\\
    p \cdot \qtau & = \frac{(1 - \beta_\tau)(M_B^2 + q^2 - M_D^2) - \beta_\tau \sqrt{\lambda} \cos\thetatau}{2}\\
    p \cdot \qmu  & = \frac{1}{2} \beta_{\nu\bar\nu} \Big[(M_B^2 + q^2 - M_D^2)((1 - \beta_\tau) + \beta_\tau\cos\thetast\\
\nonumber         & \quad - \sqrt{\lambda}(\beta_\tau + (1 - \beta_\tau) \cos\thetast) \cos\thetatau\\
\nonumber         & \quad + \sqrt{\lambda} \sqrt{\frac{1}{2} - \beta_\tau}\sin\thetast \sin\theta_\tau \cos\phi\Big]\,.
\end{align}

The scalar product involving $q$ read
\begin{align}
    q \cdot \qtau     & = (1 - \beta_\tau) q^2\,,\\
    q \cdot \qmu      & = \beta_{\nu\bar\nu} ((1 - \beta_\tau) + \beta_\tau \cos\thetast) q^2\,,\\
    q \cdot \qnubarmu & = \frac{1}{2} \Big[(1 - \beta_{\nu\bar\nu})(1 - \beta_\tau) - \beta_{\nu\bar\nu} (1 - \beta_\tau) \cos\thetastst\\
\nonumber             & \quad - \beta_\tau (\beta_{\nu\bar\nu} - (1 - \beta_{\nu\bar\nu}) \cos\thetastst)\cos\thetast\\
\nonumber             & \quad - 2 \sqrt{\frac{1}{2} - \beta_{\nu\bar\nu}} \beta_\tau \sin\thetast \sin\thetastst \cos\phistst\Big] q^2\,.
\end{align}

For scalar products involving $\qtau$ we find
\begin{align}
    \qtau \cdot \qmu      & = \beta_{\nu\bar\nu} \qtau^2\,,\\
    \qtau \cdot \qnubarmu & = \frac{1}{2}\big[(1 - \beta_{\nu\bar\nu}) - \beta_{\nu\bar\nu} \cos\thetastst\big] \qtau^2\,.
\end{align}

For the antisymmetric tensors we obtain
\begin{align}
    \eps(p, q, \qmu, \qnubarmu) & = \frac{\beta_{\nu\bar\nu} \beta_\tau \sqrt{\frac{1}{2} - \beta_\tau}}{2} \sqrt{\lambda} q^2 \sin\thetast \sin\thetatau \sin \phi\,,
\end{align}

In all of the above, we abbreviate
\begin{align}
    \beta_\tau         & = \frac{q^2 + \qtau^2}{2 q^2}\,, &
    \beta_{\nu\bar\nu} & = \frac{\qtau^2 + \qnunubar^2}{2 \qtau^2}\,. &
\end{align}

\section{Results for the Legendre Ansatz in $P_3(y)$}
\label{app:pdf-results}

The mean values and covariance matrices for the Legendre moments in the
PDFs $P_3(y)$ of $\bar{B}\to D\tau(\to \mu\bar{\nu}\nu)\bar{\nu}$ and
$\bar{B}\to \pi\tau(\to \mu\bar{\nu}\nu)\bar{\nu}$ decays are listed in
tables \ref{tab:results-legendre-moments-BtoD} and \ref{tab:results-legendre-moments-Btopi},
respectively.

\newcolumntype{d}{D{.}{.}{8}}
\begin{turnpage}
\begin{table}
\renewcommand{\arraystretch}{1.4}
\begin{tabular}{r|dddddddddddd}
    $c_k$
    & -5.02\times 10^{-1} & -4.82\times 10^{-1} &  7.41\times 10^{-1}
    & -1.98\times 10^{-1} & -1.71\times 10^{-1} &  1.85\times 10^{-1}
    & -9.99\times 10^{-2} &  1.99\times 10^{-2} &  2.67\times 10^{-2}
    & -3.22\times 10^{-2} &  1.45\times 10^{-2} & 2.46\times 10^{-3}\\
\hline
    $k$ & 1 & 2 & 3 & 4 & 5 & 6 & 7 & 8 & 9 & 10 & 11 & 12\\
\hline
    1 & 5.22\times 10^{-7} & -6.69\times 10^{-7} & -4.86\times 10^{-7} & 1.05\times 10^{-6} & -3.07\times 10^{-7} & -2.92\times 10^{-7} & 2.84\times 10^{-7} & -1.28\times 10^{-7} & 1.54\times 10^{-8} & 3.84\times 10^{-8} & -4.05\times 10^{-8} & 1.27\times 10^{-8} \\
    2 & -6.69\times 10^{-7} & 1.33\times 10^{-6} & -7.86\times 10^{-8} & -1.75\times 10^{-6} & 1.52\times 10^{-6} & -5.29\times 10^{-9} & -6.93\times 10^{-7} & 4.86\times 10^{-7} & -1.59\times 10^{-7} & -3.22\times 10^{-8} & 8.67\times 10^{-8} & -5.88\times 10^{-8} \\
    3 & -4.86\times 10^{-7} & -7.86\times 10^{-8} & 1.94\times 10^{-6} & -1.11\times 10^{-6} & -1.74\times 10^{-6} & 2.05\times 10^{-6} & -2.31\times 10^{-7} & -7.24\times 10^{-7} & 5.56\times 10^{-7} & -2.21\times 10^{-7} & 3.3\times 10^{-9} & 8.87\times 10^{-8} \\
    4 & 1.05\times 10^{-6} & -1.75\times 10^{-6} & -1.11\times 10^{-6} & 4.09\times 10^{-6} & -1.78\times 10^{-6} & -2.56\times 10^{-6} & 2.89\times 10^{-6} & -4.67\times 10^{-7} & -8.47\times 10^{-7} & 7.25\times 10^{-7} & -3.16\times 10^{-7} & 2.88\times 10^{-8} \\
    5 & -3.07\times 10^{-7} & 1.52\times 10^{-6} & -1.74\times 10^{-6} & -1.78\times 10^{-6} & 4.94\times 10^{-6} & -1.94\times 10^{-6} & -3.15\times 10^{-6} & 3.43\times 10^{-6} & -5.29\times 10^{-7} & -9.97\times 10^{-7} & 8.51\times 10^{-7} & -3.7\times 10^{-7} \\
    6 & -2.92\times 10^{-7} & -5.29\times 10^{-9} & 2.05\times 10^{-6} & -2.56\times 10^{-6} & -1.94\times 10^{-6} & 5.89\times 10^{-6} & -2.37\times 10^{-6} & -3.58\times 10^{-6} & 3.93\times 10^{-6} & -6.18\times 10^{-7} & -1.12\times 10^{-6} & 9.4\times 10^{-7} \\
    7 & 2.84\times 10^{-7} & -6.93\times 10^{-7} & -2.31\times 10^{-7} & 2.89\times 10^{-6} & -3.15\times 10^{-6} & -2.37\times 10^{-6} & 6.9\times 10^{-6} & -2.75\times 10^{-6} & -4.06\times 10^{-6} & 4.45\times 10^{-6} & -7.26\times 10^{-7} & -1.23\times 10^{-6} \\
    8 & -1.28\times 10^{-7} & 4.86\times 10^{-7} & -7.24\times 10^{-7} & -4.67\times 10^{-7} & 3.43\times 10^{-6} & -3.58\times 10^{-6} & -2.75\times 10^{-6} & 7.84\times 10^{-6} & -3.1\times 10^{-6} & -4.57\times 10^{-6} & 4.97\times 10^{-6} & -8.\times 10^{-7} \\
    9 & 1.54\times 10^{-8} & -1.59\times 10^{-7} & 5.56\times 10^{-7} & -8.47\times 10^{-7} & -5.29\times 10^{-7} & 3.93\times 10^{-6} & -4.06\times 10^{-6} & -3.1\times 10^{-6} & 8.75\times 10^{-6} & -3.44\times 10^{-6} & -5.04\times 10^{-6} & 5.46\times 10^{-6} \\
    0 & 3.84\times 10^{-8} & -3.22\times 10^{-8} & -2.21\times 10^{-7} & 7.25\times 10^{-7} & -9.97\times 10^{-7} & -6.18\times 10^{-7} & 4.45\times 10^{-6} & -4.57\times 10^{-6} & -3.44\times 10^{-6} & 9.68\times 10^{-6} & -3.81\times 10^{-6} & -5.51\times 10^{-6} \\
    11 & -4.05\times 10^{-8} & 8.67\times 10^{-8} & 3.3\times 10^{-9} & -3.16\times 10^{-7} & 8.51\times 10^{-7} & -1.12\times 10^{-6} & -7.26\times 10^{-7} & 4.97\times 10^{-6} & -5.04\times 10^{-6} & -3.81\times 10^{-6} & 1.06\times 10^{-5} & -4.14\times 10^{-6} \\
    12 & 1.27\times 10^{-8} & -5.88\times 10^{-8} & 8.87\times 10^{-8} & 2.88\times 10^{-8} & -3.7\times 10^{-7} & 9.4\times 10^{-7} & -1.23\times 10^{-6} & -8.\times 10^{-7} & 5.46\times 10^{-6} & -5.51\times 10^{-6} & -4.14\times 10^{-6} & 1.15\times 10^{-5} \\
\end{tabular}
\renewcommand{\arraystretch}{1.0}
\caption{
    \label{tab:results-legendre-moments-BtoD}
    Mean values and covariance matrix for the Legendre moments $c^{(3)}_k$ in the parametrization of the
    PDF $P_3(y)$ in the Decay $\bar{B}\to D \tau(\to \mu \bar{\nu}\nu)\bar\nu$. We use $k \leq 12$.
}
\end{table}

\begin{table}
\renewcommand{\arraystretch}{1.4}
\begin{tabular}{r|dddddddddddd}
    $c_k$
    & -5.2\times 10^{-1}  & -3.6\times 10^{-1}  &  6.14\times 10^{-1}
    & -3.02\times 10^{-1} &  6.0\times 10^{-2}  &  6.47\times 10^{-2}
    & -1.18\times 10^{-1} &  1.09\times 10^{-1} & -6.88\times 10^{-2}
    & 2.31\times 10^{-2}  & 1.11\times 10^{-2}  & -3.\times 10^{-2} \\
\hline
    $k$ & 1 & 2 & 3 & 4 & 5 & 6 & 7 & 8 & 9 & 10 & 11 & 12\\
\hline
    1 & 6.58\times 10^{-7} & -7.83\times 10^{-7} & -5.16\times 10^{-7} & 1.03\times 10^{-6} & -4.39\times 10^{-7} & -4.1\times 10^{-12} & 1.7\times 10^{-7} & -2.17\times 10^{-7} & 1.69\times 10^{-7} & -8.57\times 10^{-8} & 5.91\times 10^{-9} & 3.7\times 10^{-8} \\
    2 & -7.83\times 10^{-7} & 1.62\times 10^{-6} & -2.61\times 10^{-7} & -1.79\times 10^{-6} & 1.74\times 10^{-6} & -4.72\times 10^{-7} & -2.72\times 10^{-7} & 4.94\times 10^{-7} & -4.59\times 10^{-7} & 2.78\times 10^{-7} & -8.51\times 10^{-8} & -6.28\times 10^{-8} \\
    3 & 5.16\times 10^{-7} & -2.61\times 10^{-7} & 2.24\times 10^{-6} & -1.04\times 10^{-6} & -1.83\times 10^{-6} & 2.\times 10^{-6} & -5.54\times 10^{-7} & -2.55\times 10^{-7} & 4.78\times 10^{-7} & -4.5\times 10^{-7} & 2.7\times 10^{-7} & -7.25\times 10^{-8} \\
    4 & 1.03\times 10^{-6} & -1.79\times 10^{-6} & -1.04\times 10^{-6} & 4.14\times 10^{-6} & -1.93\times 10^{-6} & -2.14\times 10^{-6} & 2.58\times 10^{-6} & -9.36\times 10^{-7} & -6.37\times 10^{-8} & 4.21\times 10^{-7} & -4.79\times 10^{-7} & 3.73\times 10^{-7} \\
    5 & -4.39\times 10^{-7} & 1.74\times 10^{-6} & -1.83\times 10^{-6} & -1.93\times 10^{-6} & 5.3\times 10^{-6} & -2.28\times 10^{-6} & -2.71\times 10^{-6} & 3.21\times 10^{-6} & -1.22\times 10^{-6} & -9.6\times 10^{-9} & 5.19\times 10^{-7} & -6.59\times 10^{-7} \\
    6 & -4.1\times 10^{-12} & -4.72\times 10^{-7} & 2.\times 10^{-6} & -2.14\times 10^{-6} & -2.28\times 10^{-6} & 6.14\times 10^{-6} & -2.55\times 10^{-6} & -3.21\times 10^{-6} & 3.72\times 10^{-6} & -1.35\times 10^{-6} & -1.12\times 10^{-7} & 6.54\times 10^{-7} \\
    7 & 1.7\times 10^{-7} & -2.72\times 10^{-7} & -5.54\times 10^{-7} & 2.58\times 10^{-6} & -2.71\times 10^{-6} & -2.55\times 10^{-6} & 7.03\times 10^{-6} & -2.91\times 10^{-6} & -3.57\times 10^{-6} & 4.09\times 10^{-6} & -1.45\times 10^{-6} & -1.11\times 10^{-7} \\
    8 & -2.17\times 10^{-7} & 4.94\times 10^{-7} & -2.55\times 10^{-7} & -9.36\times 10^{-7} & 3.21\times 10^{-6} & -3.21\times 10^{-6} & -2.91\times 10^{-6} & 8.05\times 10^{-6} & -3.41\times 10^{-6} & -3.92\times 10^{-6} & 4.56\times 10^{-6} & -1.68\times 10^{-6} \\
    9 & 1.69\times 10^{-7} & -4.59\times 10^{-7} & 4.78\times 10^{-7} & -6.37\times 10^{-8} & -1.22\times 10^{-6} & 3.72\times 10^{-6} & -3.57\times 10^{-6} & -3.41\times 10^{-6} & 9.09\times 10^{-6} & -3.79\times 10^{-6} & -4.39\times 10^{-6} & 5.1\times 10^{-6} \\
    10 & -8.57\times 10^{-8} & 2.78\times 10^{-7} & -4.5\times 10^{-7} & 4.21\times 10^{-7} & -9.6\times 10^{-9} & -1.35\times 10^{-6} & 4.09\times 10^{-6} & -3.92\times 10^{-6} & -3.79\times 10^{-6} & 9.99\times 10^{-6} & -4.11\times 10^{-6} & -4.83\times 10^{-6} \\
    11 & 5.91\times 10^{-9} & -8.51\times 10^{-8} & 2.7\times 10^{-7} & -4.79\times 10^{-7} & 5.19\times 10^{-7} & -1.12\times 10^{-7} & -1.45\times 10^{-6} & 4.56\times 10^{-6} & -4.39\times 10^{-6} & -4.11\times 10^{-6} & 1.09\times 10^{-5} & -4.5\times 10^{-6} \\
    12 & 3.7\times 10^{-8} & -6.28\times 10^{-8} & -7.25\times 10^{-8} & 3.73\times 10^{-7} & -6.59\times 10^{-7} & 6.54\times 10^{-7} & -1.11\times 10^{-7} & -1.68\times 10^{-6} & 5.1\times 10^{-6} & -4.83\times 10^{-6} & -4.5\times 10^{-6} & 1.19\times 10^{-5} \\
\end{tabular}
\renewcommand{\arraystretch}{1.0}
\caption{
    \label{tab:results-legendre-moments-Btopi}
    Mean values and covariance matrix for the Legendre moments $c^{(3)}_k$ in the parametrization of the
    PDF $P_3(y)$ in the Decay $\bar{B}\to \pi \tau(\to \mu \bar{\nu}\nu)\bar\nu$.
    We use $k \leq 12$.
}
\end{table}
\end{turnpage}

\clearpage

\bibliography{references}

\end{document}